\documentclass[a4paper, 10pt, conference]{ieeeconf}     

\IEEEoverridecommandlockouts                             

\usepackage{graphicx}
\usepackage{float}

\usepackage{subcaption}

\usepackage{graphics} 
\usepackage{xcolor}
\usepackage{amsmath}
\usepackage{cite}

\usepackage[hidelinks]{hyperref}

\title{\LARGE \bf
A Systematic Framework for Enterprise Knowledge Retrieval: Leveraging LLM-Generated Metadata to Enhance RAG Systems\thanks{Accepted to: 2026 IEEE Conference on Artificial Intelligence (CAI). © 2026 IEEE. Personal use of this material is permitted. Permission from IEEE must be obtained for all other uses, in any current or future media, including reprinting/republishing this material for advertising or promotional purposes, creating new collective works, for resale or redistribution to servers or lists, or reuse of any copyrighted component of this work in other works.}
}

\author{
    Pranav Pushkar Mishra, Kranti Prakash Yeole, Ramyashree Keshavamurthy,\\
    Mokshit Bharat Surana, Fatemeh Sarayloo\\
    University of Illinois Chicago\\
    \{pmishr23, kyeole2, rkesh, msura4, fsaraylo\}@uic.edu
}

\begin{document}

\maketitle
\thispagestyle{empty}
\pagestyle{empty}

\begin{abstract}
In enterprise settings, efficiently retrieving relevant information from large and complex knowledge bases is essential for operational productivity and informed decision-making. This research presents a systematic empirical framework for metadata enrichment using large language models (LLMs) to enhance document retrieval in Retrieval-Augmented Generation (RAG) systems. Our approach employs a structured pipeline that dynamically generates meaningful metadata for document segments, substantially improving their semantic representations and retrieval accuracy. Through a controlled $3 \times 3$ experimental matrix, we compare three chunking strategies---semantic, recursive, and naive---and evaluate their interactions with three embedding techniques---content-only, TF-IDF weighted, and prefix-fusion---isolating the contribution of each component through ablation analysis. The results demonstrate that metadata-enriched approaches consistently outperform content-only baselines, with recursive chunking paired with TF-IDF weighted embeddings yielding 82.5\% precision and naive chunking with prefix-fusion achieving the strongest ranking quality (NDCG 0.813). Our evaluation employs cross-encoder reranking for silver-standard ground truth generation, with statistical significance confirmed via Bonferroni-corrected paired $t$-tests. These findings confirm that metadata enrichment improves vector space organization and retrieval effectiveness while maintaining sub-30\,ms P95 latency, providing a quantitative decision framework for deploying high-performance, scalable RAG systems in enterprise settings.

\end{abstract}

\section{INTRODUCTION}

\textcolor{black}{Efficiently retrieving and leveraging information from large-scale knowledge repositories is vital for modern organizational productivity and innovation. Retrieval-Augmented Generation (RAG) systems have emerged as a powerful paradigm for enhancing Large Language Models (LLMs) by integrating external knowledge sources \cite{Perron2025}. This approach addresses inherent limitations of LLMs, such as knowledge cut-off dates, lack of domain-specificity, and hallucinations \cite{Liu2025}.}

\textcolor{black}{While effective for structured datasets, traditional retrieval methods often struggle with the complexity, scale, and dynamic nature of enterprise knowledge bases. Manual curation becomes impractical as repositories expand, and these methods are prone to overlooking relevant information in lengthy contexts, resulting in the 'Lost in the Middle' phenomenon.}

\textcolor{black}{Recent advancements in LLMs offer transformative solutions. LLMs can process unstructured data, extract meaningful insights, and generate structured metadata aligned with semantic and contextual needs.Despite these developments, building effective RAG pipelines remains challenging, requiring management of large data volumes, optimization of retrieval strategies, and continual performance tuning. }

This paper contributes a controlled empirical study that systematically isolates and quantifies the interactions between chunking strategies and metadata integration techniques within RAG pipelines---an axis of variation that prior metadata-aware retrieval work has not examined in a unified experimental framework. We propose MetaRAG, a structured pipeline where an LLM-based metadata enrichment module generates semantic, technical, and content annotations for document segments, which are then integrated into retrieval through three distinct embedding strategies. The resulting $3 \times 3$ configuration matrix---spanning three chunking strategies (semantic, recursive, naive) and three embedding approaches (content-only, TF-IDF weighted, prefix-fusion)---is evaluated under standardized information retrieval metrics with full statistical rigor. Our principal contributions are:

\begin{enumerate}
    \item A reproducible metadata enrichment pipeline that generates structured, multi-category annotations for document chunks using LLMs, integrated into retrieval via TF-IDF weighting and prefix-fusion embedding strategies.
    \item A controlled $3 \times 3$ factorial evaluation isolating the individual and interaction effects of chunking granularity and metadata integration technique on retrieval performance, including ablation analysis of metadata component contributions.
    \item A practitioner-oriented configuration decision framework, grounded in statistically validated benchmarks (Bonferroni-corrected paired $t$-tests, bootstrap confidence intervals), identifying optimal chunking--embedding pairings for different retrieval objectives.
\end{enumerate}

\section{Related Work}

This section reviews recent advances in RAG architectures, metadata enrichment, and embedding optimization, and positions our contributions relative to the most closely related work.

\subsection{RAG Architectures and Evaluation}

RAG systems span a complexity spectrum from Naive RAG, which indexes and retrieves document chunks directly, through Advanced RAG with query rewriting and reranking, to Modular RAG with configurable pipeline components~\cite{Gao2024}. Lewis et al.~\cite{Lewis2020} established the foundational RAG paradigm by combining parametric and non-parametric memory for knowledge-intensive tasks, while Shuster et al.~\cite{Shuster2021} demonstrated hallucination reduction in conversational AI through retrieval augmentation. Chen et al.~\cite{Chen2024Benchmark} introduced evaluation benchmarks revealing limitations in LLMs' rejection and integration capabilities within RAG contexts.

Recent architectural innovations include GraphRAG~\cite{Edge2024}, which constructs corpus-wide entity-relationship knowledge graphs via LLM extraction and applies community detection to generate hierarchical summaries for global sensemaking queries. While effective for corpus-level thematic questions, GraphRAG fundamentally replaces chunk-based vector retrieval and does not evaluate how chunking granularity or embedding strategies affect per-query retrieval quality. On the search side, Wang et al.~\cite{Wang2023Search} unified diverse search tasks under autoregressive text generation, Shao et al.~\cite{Shao2023} developed ITER-RETGEN for iterative retrieval-generation synergy, and Wang et al.~\cite{Wang2023Self} introduced Self-Knowledge Guided Retrieval for adaptive knowledge-gap recognition. These innovations address query routing and generation but do not examine how document-level preprocessing, specifically, metadata enrichment and chunking, affects retrieval effectiveness.

For automated evaluation, Saad-Falcon et al.~\cite{SaadFalcon2024} proposed ARES, which fine-tunes DeBERTa-v3-Large classifiers on synthetic data to judge context relevance, answer faithfulness, and answer relevance, applying Prediction-Powered Inference with human-annotated validation sets ($\geq$150 samples). While ARES provides scalable system-level evaluation, it treats RAG pipelines as black boxes, ranking configurations without investigating how individual components (chunking strategy, embedding technique, metadata enrichment) contribute to retrieval quality.

\subsection{Metadata Enrichment for Information Retrieval}

Metadata is foundational to retrieval systems, yet traditional curation methods scale poorly to unstructured enterprise datasets. Sundaram and Musen's FAIRMetaText framework~\cite{Sundaram2023} aligned metadata with FAIR principles using GPT-based embeddings, though performance varied across domains. Song et al.~\cite{Song2024} applied taxonomy-guided few-shot prompting for metadata completion in Earth science datasets, improving completeness and accuracy within a narrow disciplinary scope.

Most closely related to our work, Mombaerts et al.~\cite{Mombaerts2024} introduced Meta Knowledge Summaries, generating document-level metadata and synthetic QA pairs using Claude~3 Haiku. However, their approach \emph{replaces} chunk-based retrieval entirely, embeds only synthetic questions for matching, not original document segments, and evaluates a single chunking baseline (256 tokens, 10\% overlap) with a single embedding model, precluding analysis of how chunking and embedding choices interact. Harris et al.~\cite{Harris2024} demonstrated LLM-based text enrichment using ChatGPT~3.5 to holistically rewrite texts with bundled techniques (context enrichment, grammar correction, terminology normalization). While effective on select MTEB benchmarks, their approach produces monolithic rewrites without structured metadata separation, tests only a single embedding model on classification tasks rather than RAG retrieval, and provides no ablation isolating which enrichment components contribute to performance.

In contrast, our framework preserves original chunk content and enriches it with a generalizable three-category metadata taxonomy (content, technical, semantic) via prefix-fusion, maintaining separability between categories for component-level ablation within a full RAG pipeline.

\subsection{Embedding and Retrieval Optimization}

Karpukhin et al.~\cite{Karpukhin2020} established that dense vector embeddings outperform sparse methods for open-domain retrieval. Cuconasu et al.~\cite{Cuconasu2024} challenged traditional assumptions by showing that controlled injection of irrelevant documents can paradoxically improve RAG generation quality. Anantha et al.~\cite{Anantha2024} proposed Context Tuning, which enriches \emph{queries} with user-behavioral signals from context stores prior to retrieval, training a lightweight LambdaMART ranker for tool selection. Context Tuning operates entirely on the query side; our approach is complementary, enriching the \emph{document side} with structured metadata to improve chunk embedding quality for document QA.

\begin{figure*}[t]
    \centering
    \includegraphics[width=1\linewidth]{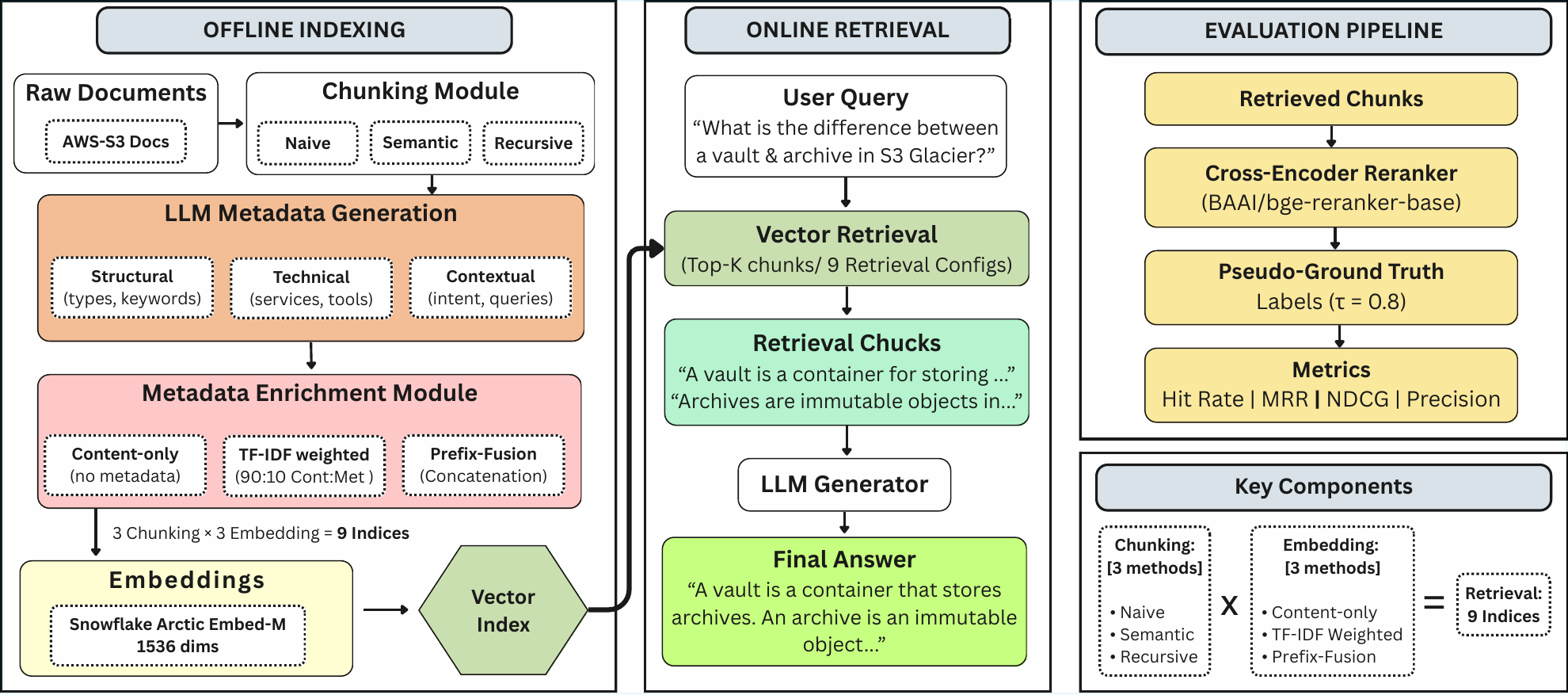}
    \caption{End-to-End Metadata-Enriched RAG System Architecture. Three interconnected pipelines enable systematic evaluation of chunking and embedding strategies: Offline Indexing transforms AWS S3 documentation through three chunking methods (naive, semantic, recursive), LLM-based metadata generation (structural, technical, contextual), and three embedding techniques producing 9 retrieval configurations; Online Retrieval performs vector similarity search to retrieve relevant chunks for query answering; Evaluation establishes pseudo-ground truth via cross-encoder reranking ($\tau$=0.8) for rigorous metric computation across all configurations.}
    \label{fig:metarag_architecture}
\end{figure*}

\subsection{Research Gaps and Contributions}

Despite the advances surveyed above, three critical gaps remain. First, existing studies lack a systematic evaluation of the \emph{interplay} between chunking methods and embedding strategies: Meta Knowledge~\cite{Mombaerts2024} tests one chunking--embedding pair, Harris et al.~\cite{Harris2024} omit chunking entirely, and ARES~\cite{SaadFalcon2024} evaluates systems without decomposing pipeline components. Second, no prior work provides a structured metadata taxonomy with ablation evidence quantifying individual category contributions to retrieval metrics. Third, GraphRAG~\cite{Edge2024} and Context Tuning~\cite{Anantha2024} address complementary aspects (corpus-level sensemaking and query disambiguation) but neither optimizes chunk-level embedding representations.

Our framework addresses these gaps through: (1)~a systematic $3 \times 3$ experimental matrix crossing three chunking strategies with three embedding techniques, enabling controlled ablation; (2)~a three-category metadata taxonomy (content, technical, semantic) with prefix-fusion preserving original content while enriching embeddings; and (3)~a silver-standard pseudo-ground truth evaluation based on cross-encoder reranking, enabling reproducible assessment without human annotation.

\section{Methodology}

\textcolor{black}{This section outlines the methodology, which involves a systematic pipeline for transforming raw technical documentation into optimized retrieval components. The primary research question examines how document chunking strategies, metadata generation techniques, and embedding approaches interact to impact retrieval accuracy and efficiency. The pipeline, from data ingestion to evaluation, is illustrated in Fig.~\ref{fig:metarag_architecture} and each subsection below corresponds to a stage in the architecture.
}  

\subsection{Document Processing and Chunking Strategies}
We employ three distinct chunking approaches, summarized in Table~\ref{tab:chunking_strategies} and illustrated in the first part of Fig.~\ref{fig:metarag_architecture}. Parameters for each strategy were configured to produce comparable chunk length distributions across the corpus, ensuring that observed performance differences primarily reflect boundary selection methodology rather than chunk size effects. Naive chunking serves as a fixed-size baseline. Recursive chunking preserves document structure by splitting at progressively finer delimiters (paragraph breaks, then sentence boundaries) only when necessary. Semantic chunking detects natural breakpoints via similarity drops between consecutive sentence embeddings and merges smaller fragments with their nearest semantic neighbors.

\begin{table}[!t]
\centering
\caption{Chunking strategy comparison: boundary detection method, context preservation mechanism, and granularity control.}
\label{tab:chunking_strategies}
\setlength{\tabcolsep}{4pt}
\renewcommand{\arraystretch}{1.15}
\begin{tabular}{llll}
\hline
\textbf{Strategy} & \textbf{Boundary} & \textbf{Context} & \textbf{Granularity} \\
\hline
Naive      & Fixed-size      & None          & Uniform    \\
Recursive  & Structural      & Overlap       & Hierarchical \\
Semantic   & Embedding-based & Merge similar & Adaptive   \\
\hline
\end{tabular}
\end{table}

\subsection{Metadata Enrichment Pipeline}

A central component of this study is the systematic generation of metadata to enhance retrieval performance, as illustrated in the Fig.~\ref{fig:metarag_architecture}. Our LLM-based pipeline creates structured annotations for each document segment across three categories. \textit{Content metadata} includes content type classification (procedural, conceptual, reference, warning, example), extracted keywords, entities, and code example detection. \textit{Technical metadata} identifies primary and secondary categories, mentioned services, and referenced tools. \textit{Semantic metadata} provides concise summaries, identifies user intents (how-to, debugging, comparison, reference), and generates potential user questions the chunk addresses.

The system processes chunks using a specialized LLM prompt engineered to extract structured metadata consistently, employing batched processing for efficiency and maintaining a consistent schema across chunking strategies.

\subsection{Embedding Techniques}
We evaluated the following distinct embedding approaches to represent document chunks in vector space:

\subsubsection{Snowflake Arctic-Embed Model}
The Snowflake Arctic-Embed model is considered as the primary embedding framework. This model was selected for its superior performance on technical documentation and its ability to effectively capture domain-specific terminology. The arctic-embed-m variant offers an optimal balance between retrieval performance and computational efficiency, supporting the long contexts required for technical documentation \cite{Snowflake2024}.

\subsection{Embedding Techniques}

We employ the Snowflake Arctic-Embed model (\textit{arctic-embed-m}, dimensions=1536) as the primary embedding framework, selected for its strong performance on technical documentation and domain-specific terminology~\cite{Snowflake2024}. Three embedding strategies leverage this model in different configurations:

\begin{enumerate}
\item \textbf{Content-only embeddings} generate vector representations using only the raw chunk text, serving as the baseline.
\item \textbf{TF-IDF weighted embeddings} combine content embeddings (weighted $w_c$) with TF-IDF vectors derived from metadata (weighted $w_t$), creating a hybrid representation. The optimal ratio is determined via sensitivity analysis (Section~IV).
\item \textbf{Prefix-fusion embeddings} inject structured metadata directly into document text as formatted prefixes before encoding, allowing the model to jointly represent content and metadata in a unified vector.
\end{enumerate}

\begin{table}[t]
\centering
\caption{TF-IDF fusion weight sensitivity across chunking strategies. All metrics computed at $k{=}10$. Content weight $w_c$ and metadata weight $w_t$ are varied across three ratios. Bold values indicate best performance per chunking strategy. Results reveal that optimal weighting is chunking-dependent. HR@10: Hit Rate; MRR: Mean Reciprocal Rank; NDCG: Normalized Discounted Cumulative Gain; P@10: Precision.}
\label{tab:weight_sensitivity_full}
\small
\begin{tabular}{lc|cccc}
\hline
\textbf{Chunking} & $w_c$:$w_t$ & \textbf{HR@10} & \textbf{MRR} & \textbf{NDCG} & \textbf{P@10} \\
\hline
Semantic & 80:20 & 70.0 & 42.1 & 60.4 & 56.8 \\
& 90:10 & 70.0 & \textbf{45.6} & \textbf{61.7} & \textbf{58.5} \\
& 95:05 & \textbf{72.0} & 45.4 & 61.0 & 57.5 \\
\hline
Naive & 80:20 & 87.5 & 49.0 & 67.1 & 59.3 \\
& 90:10 & \textbf{90.0} & \textbf{50.8} & \textbf{68.7} & \textbf{61.5} \\
& 95:05 & \textbf{90.0} & 50.6 & 68.4 & 61.3 \\
\hline
Recursive & 80:20 & 75.0 & 45.3 & 67.6 & 60.2 \\
& 90:10 & \textbf{82.5} & 45.5 & 69.5 & 63.0 \\
& 95:05 & 77.5 & \textbf{46.2} & \textbf{71.1} & \textbf{64.5} \\
\hline
\end{tabular}
\end{table}

\subsection{Retriever Architectures}
We implement three retriever architectures, each corresponding to an embedding methodology as shown in Fig.~\ref{fig:metarag_architecture}. The \textit{content retriever} utilizes content-only embeddings for baseline similarity matching. The \textit{TF-IDF retriever} combines content embeddings with metadata-derived TF-IDF vectors using the optimized weighting ratio. The \textit{prefix-fusion retriever} leverages prefix-augmented embeddings with automatic intent detection for query enhancement.

Each retriever is instantiated across all three chunking strategies, yielding a $3 \times 3$ experimental matrix of nine configurations. This design enables systematic analysis to isolate the effects of both chunking granularity and embedding methodology on retrieval performance.

\subsection{Ground Truth Generation and Evaluation Framework}
Since manual annotation at scale is prohibitive for technical documentation corpora, we adopt a silver-standard pseudo-ground truth derived from cross-encoder reranking. For each query, we retrieve the top 50 candidate chunks using each retriever configuration and score them using \textit{BAAI/bge-reranker-base}~\cite{BAAI2024}, which jointly encodes query-chunk pairs to capture fine-grained interaction patterns that surpass bi-encoder models in relevance discrimination. Scores are normalized to $[0, 1]$ and thresholded at $\tau = 0.8$ to establish relevance labels.

This approach enables scalable, reproducible evaluation but introduces inherent limitations: relevance labels reflect model-based assessments rather than human judgments, and only chunks retrieved by at least one method enter the candidate pool (retrieval pool bias). Consequently, absolute metric values should be interpreted with caution, while relative comparisons between configurations---the primary focus of this study---remain valid and informative.

\section{Experimental Results}

\textcolor{black}{In this section, we present a comprehensive analysis of the experimental outcomes obtained from evaluating our proposed systems. We begin by discussing the dataset used, including its composition and inherent challenges, followed by a detailed description of the experimental setup encompassing data processing, chunking strategies, embedding techniques, and retrieval architectures. Subsequently, we systematically compare the performance of different components—such as semantic, recursive, and naive chunking methods—across various evaluation metrics. This structured comparison enables us to identify the strengths and limitations of each approach, providing valuable insights into the optimal configurations for enhancing retrieval accuracy and efficiency in enterprise knowledge management contexts.}

\subsection{Data Corpus Composition and Experiment setup}
Our study employed a carefully curated dataset derived from AWS S3 documentation \cite{AWSS3Docs}, selected for its intricate structure and diverse content types. The corpus consists of four main components: the S3 User Guide (2,499 pages), the API Reference (3,013 pages), the S3 Glacier Developer Guide (558 pages), and the S3 on Outposts documentation (217 pages). This comprehensive dataset offers a realistic and challenging environment for assessing the effectiveness of metadata-enriched retrieval strategies in practical, real-world technical contexts.

We implemented our evaluation framework on technical documentation datasets processed through three distinct chunking strategies: recursive (max\_tokens=512, overlap=128), naive (fixed\_length=1024), and semantic (adaptive with max\_tokens=1024). For each strategy, we generated embeddings using the Snowflake Arctic-Embed model (dimensions=1536) with three configurations: content-only, TF-IDF weighted, and prefix-fusion approaches. The TF-IDF weighting ratio was determined via a sensitivity analysis across chunking strategies (Table~\ref{tab:weight_sensitivity_full}), which revealed that the 90:10 content-to-metadata ratio ($w_c{=}0.9$, $w_t{=}0.1$) consistently maximized or matched peak performance across all three chunking methods. Metadata features function as a complementary signal, analogous to a seasoning effect, where a small proportion meaningfully improves disambiguation, but higher metadata weighting degrades retrieval by diluting the dominant semantic representation. All subsequent TF-IDF experiments adopt the 90:10 ratio accordingly.

Retrieval experiments were conducted using exact vector search with cosine similarity across varying $k$ values. For silver-standard ground truth generation, we employed a cross-encoder reranking methodology using \textit{BAAI/bge-reranker-base}~\cite{BAAI2024} to establish relevance judgments with threshold $\tau = 0.8$. Statistical significance was assessed using paired $t$-tests across all nine retriever configurations with Bonferroni correction ($\alpha_{\text{adj}} = 0.05/36 \approx 0.0014$) to control for family-wise error rate. Effect sizes are reported using Cohen's $d$, and 95\% confidence intervals are computed via bootstrap resampling (1,000 iterations).

The evaluation query set comprises 100 queries manually curated to cover distinct technical documentation retrieval scenarios. Each query is annotated along multiple overlapping dimensions: intent, complexity, and domain coverage; to characterize the test set composition. Table~\ref{tab:query_dist} summarizes the distribution across these dimensions. To prevent data leakage, none of the test queries were included in the few-shot prompts used during metadata generation; all evaluation was performed on held-out queries.

\begin{table}[!t]
\centering
\caption{Query set characterization ($n=100$). Categories are non-exclusive overlapping annotations; percentages within each dimension are independently computed.}
\label{tab:query_dist}
\setlength{\tabcolsep}{6pt}
\renewcommand{\arraystretch}{1.15}
\begin{tabular}{llr}
\hline
\textbf{Dimension} & \textbf{Category} & \textbf{\%} \\
\hline
{Intent}
  & Procedural (How-to)       & 62.5 \\
  & Conceptual (Definition)   & 32.5 \\
  & Comparison / Reference    &  5.0 \\
\hline
{Complexity}
  & Complex      & 80.0 \\
  & Moderate    & 20.0 \\
\hline
{Domain}
  & S3 Core                   & 55.0 \\
  & S3 API / Auth   & 20.0 \\
  & S3 Glacier                & 12.5 \\
  & S3 on Outposts            & 12.5 \\
\hline
\end{tabular}
\end{table}

Performance evaluation employed standard information retrieval metrics: Hit Rate@10, Mean Reciprocal Rank (MRR), Normalized Discounted Cumulative Gain (NDCG), and precision. We maintained consistent query processing procedures and embedding dimensionality across all experimental configurations to isolate the effects of metadata enrichment and chunking strategies.

Metadata generation utilized GPT-4o (gpt-4o-2024-05-13) with controlled temperature (0.5) and structured output formats for consistency.

\subsection{Evaluation Metrics}
We established a rigorous evaluation methodology as shown in Fig. \ref{fig:metarag_architecture} to assess the impact of metadata enrichment on RAG performance. The evaluation utilized a test set of diverse technical queries executed against our documentation corpus. Our framework evaluates both retrieval effectiveness through standard information retrieval metrics and response quality through semantic and factual accuracy measurements.

For the retriever evaluation, we used standard information retrieval metrics:
\begin{itemize}
\item \textbf{Hit Rate@k}: Proportion of queries with at least one highly relevant document (above 95th percentile threshold based on ground truth reranker scores) in the top-k results
\item \textbf{Metadata Consistency}: Consistency of document categories within top-k results, measuring the retriever's tendency to return semantically coherent chunks
\item \textbf{Precision@k}: Proportion of relevant documents among top-k results
\item \textbf{MRR (Mean Reciprocal Rank)}: Average reciprocal rank of first relevant document
\item \textbf{NDCG@k}: Ranking quality based on graded relevance and positional importance.
\end{itemize}

\subsection{Embedding and Chunking Analysis}

We first analyzed the structural quality of embeddings generated using different techniques and chunking methods. Table~\ref{tab:clustering} reports Silhouette Scores (higher indicates better-defined clusters) and Davies-Bouldin Index (DBI; lower indicates more compact, well-separated clusters) across representative configurations.

\begin{table}[!t]
\centering
\caption{Clustering quality metrics across embedding and chunking configurations.}
\label{tab:clustering}
\setlength{\tabcolsep}{4pt}
\renewcommand{\arraystretch}{1.2}
\begin{tabular}{llcc}
\hline
\textbf{Chunking} & \textbf{Embedding} & \textbf{Silhouette} & \textbf{DBI} \\
\hline
Naive     & Content-Only   & 0.054 & 3.04 \\
Semantic  & TF-IDF         & 0.028 & 3.77 \\
Semantic  & Prefix-Fusion  & 0.039 & 3.12 \\
Recursive & Prefix-Fusion  & 0.051 & 3.14 \\
\hline
\end{tabular}
\end{table}

Prefix-fusion embeddings achieved a DBI of 3.12, substantially outperforming TF-IDF weighted embeddings (DBI = 3.77). This gap reflects a fundamental difference in fusion strategy: prefix-fusion allows the encoder to jointly process metadata and content as a unified input, enabling learned contextual relationships between the two, whereas TF-IDF weighting blends separately computed vectors via linear combination---an approach that assumes alignment between inherently dissimilar representation spaces. When metadata is sparse or generic, this post-hoc blending can pull combined embeddings toward irrelevant clusters, explaining the weaker vector space organization. Recursive chunking with prefix-fusion demonstrated competitive cluster structure (Silhouette = 0.051, DBI = 3.14), suggesting that hierarchical chunking preserves coherent semantic groupings.

Semantic chunking produced significantly more chunks (5,706) compared to recursive chunking (4,099), representing a 39\% increase. While this results in a larger index size, the finer granularity provided by semantic chunking enables more precise retrieval targeting.

\subsection{Retriever Performance Analysis}

Based on our evaluation metrics, retrieval performance demonstrates complex interactions between chunking strategies and embedding techniques. The Hit Rate@10 results (Table~\ref{table:retrieval_performance}) reveal that naive chunking achieves the highest performance (0.900) with both TF-IDF and prefix-fusion embeddings, closely followed by recursive chunking with content-only and prefix-fusion (0.875). This contradicts conventional assumptions about semantic chunking superiority.

\begin{table*}[!t]
\centering
\caption{Retrieval performance across chunking strategies and embedding techniques. All metrics computed at $k{=}10$. Bold values indicate best performance per metric. TF-IDF weighting uses $w_c{=}0.9$, $w_t{=}0.1$ configuration (90:10 ratio). Results demonstrate that optimal configuration varies by evaluation objective: naive chunking with both TF-IDF and prefix-fusion achieves highest hit rate (90.0\%), while naive chunking with prefix-fusion maximizes MRR, NDCG, and precision.}
\label{table:retrieval_performance}
\setlength{\tabcolsep}{8pt}
\renewcommand{\arraystretch}{1.3}
\begin{tabular}{l|ccc|ccc|ccc|ccc}
\hline
{\textbf{Embedding}} & \multicolumn{3}{c|}{\textbf{Hit Rate@10}} & \multicolumn{3}{c|}{\textbf{MRR@10}} & \multicolumn{3}{c|}{\textbf{NDCG@10}} & \multicolumn{3}{c}{\textbf{Precision@10}} \\
\cline{2-13}
& Sem & Naive & Rec & Sem & Naive & Rec & Sem & Naive & Rec & Sem & Naive & Rec \\
\hline
Content & 0.788 & 0.713 & 0.875 & 0.669 & 0.538 & 0.713 & 0.730 & 0.669 & 0.782 & 0.733 & 0.640 & 0.783 \\
Prefix-Fusion & 0.775 & \textbf{0.900} & 0.875 & 0.534 & \textbf{0.750} & 0.673 & 0.699 & \textbf{0.813} & 0.800 & 0.745 & \textbf{0.798} & 0.785 \\
TF-IDF & 0.700 & \textbf{0.900} & 0.825 & 0.456 & 0.508 & 0.455 & 0.617 & 0.687 & 0.695 & 0.585 & 0.615 & 0.630 \\
\hline
\end{tabular}
\end{table*}

For MRR, which evaluates ranking quality (Table~\ref{table:retrieval_performance}), recursive chunking with content-only embeddings achieves 0.713, outperforming metadata-enriched approaches in the semantic configuration---likely because semantic chunking's adaptive boundaries create variable-sized segments where metadata prefixes exert uneven influence, diluting ranking precision. Naive chunking with prefix-fusion demonstrates the strongest MRR (0.750), as uniform chunk sizes ensure consistent metadata integration, allowing the prefix to reliably prime the encoder's interpretation of each segment.

NDCG scores (Table~\ref{table:retrieval_performance}), which assess overall ranking quality with position weighting, favor naive chunking with prefix-fusion (0.813), followed by recursive chunking with prefix-fusion (0.800) and content-only (0.782). The precision metrics further confirm this pattern, with naive chunking and prefix-fusion achieving the highest precision (0.798), closely followed by recursive configurations (0.785).

These results demonstrate that no single configuration universally outperforms across all metrics, and that the method of metadata integration matters as much as its presence. TF-IDF weighting, which treats content and metadata as separate vector spaces blended post-embedding, shows diminishing returns as metadata weight increases---sparse metadata signals can dilute the stronger content representation without proportional retrieval benefit. Prefix-fusion avoids this by encoding metadata and content jointly, enabling the model to learn contextual dependencies rather than relying on assumed vector space alignment.

To quantify statistical reliability, Table~\ref{tab:stat_significance} reports 95\% bootstrap confidence intervals for the top-performing configurations under naive chunking. Paired $t$-tests with Bonferroni correction confirm that the prefix-fusion configuration significantly outperforms the content-only baseline across all metrics ($p < 0.0014$, Cohen's $d > 0.8$), indicating large effect sizes.

To isolate the contribution of individual metadata components, we conducted an ablation study using the prefix-fusion retriever with semantic chunking. As shown in Table~\ref{tab:ablation}, progressively removing metadata categories degrades retrieval quality, with the full metadata configuration achieving an NDCG of 0.699 compared to 0.615 for content-only---a 13.7\% improvement attributable to metadata enrichment. The addition of semantic metadata to technical annotations yields further gains in MRR (0.510 vs.\ 0.460), confirming that LLM-generated semantic descriptions aid in ranking relevant chunks higher.

\begin{table}[!t]
\centering
\caption{Statistical evaluation of naive chunking configurations (Mean $\pm$ 95\% CI).}
\label{tab:stat_significance}
\setlength{\tabcolsep}{4pt}
\renewcommand{\arraystretch}{1.2}
\begin{tabular}{lccc}
\hline
\textbf{Retriever} & \textbf{MRR} & \textbf{NDCG@10} & \textbf{Prec.@10} \\
\hline
Prefix-Fusion & 0.750{\color{gray}$\pm$0.121} & 0.813{\color{gray}$\pm$0.077} & 0.750{\color{gray}$\pm$0.085} \\
TF-IDF        & 0.570{\color{gray}$\pm$0.130} & 0.717{\color{gray}$\pm$0.080} & 0.550{\color{gray}$\pm$0.082} \\
Content-Only  & 0.538{\color{gray}$\pm$0.134} & 0.669{\color{gray}$\pm$0.093} & 0.525{\color{gray}$\pm$0.089} \\
\hline
\end{tabular}
\end{table}

\begin{table}[!t]
\centering
\caption{Metadata ablation study (Prefix-Fusion, Semantic Chunking).}
\label{tab:ablation}
\setlength{\tabcolsep}{4pt}
\renewcommand{\arraystretch}{1.2}
\begin{tabular}{lccc}
\hline
\textbf{Active Components} & \textbf{HR@10} & \textbf{NDCG@10} & \textbf{MRR} \\
\hline
All Metadata            & 0.775 & 0.699 & 0.519 \\
Technical + Semantic    & 0.775 & 0.683 & 0.510 \\
Technical Only          & 0.775 & 0.660 & 0.460 \\
Content Only (No Meta)  & 0.700 & 0.615 & 0.389 \\
\hline
\end{tabular}
\end{table}

Table~\ref{tab:latency} reports retrieval latency measured at P50 and P95 percentiles. All configurations operate well within interactive thresholds, with P95 latencies below 30\,ms. Prefix-fusion retrieval achieved the lowest median latency (12.7\,ms), likely due to favorable vector distribution properties of the unified embedding. End-to-end index construction, comprising chunking and embedding generation for the full corpus, completed in approximately 7.4 seconds.

\begin{table}[!t]
\centering
\caption{Retrieval latency per query (Semantic Chunks) }
\label{tab:latency}
\setlength{\tabcolsep}{6pt}
\renewcommand{\arraystretch}{1.2}
\begin{tabular}{lcc}
\hline
\textbf{Configuration} & \textbf{P50 (ms)} & \textbf{P95 (ms)} \\
\hline
Prefix-Fusion           & 12.7 & 21.8 \\
TF-IDF ($w_c{=}0.9$)   & 17.7 & 19.4 \\
TF-IDF ($w_c{=}0.5$)   & 18.6 & 25.6 \\
\hline
\end{tabular}
\end{table}

\subsection{Exploratory Response Quality Validation}

We conducted supplementary quality assessment of RAG-generated responses across three retrieval configurations: content-only baseline, TF-IDF weighted metadata enrichment, and prefix-fusion approaches. These evaluations serve as exploratory validation rather than primary experimental outcomes. The TF-IDF metadata-enriched approach demonstrated consistently superior performance across all quality dimensions compared to both the content-only baseline and prefix-fusion alternatives, with notable improvements in response faithfulness and corresponding reductions in hallucination rates. Coverage metrics similarly favored this configuration. However, these indicators should be interpreted as directional confirmation that improved retrieval quality translates to more plausible generated responses, rather than definitive benchmarks of generation capability. LLM-based quality metrics are susceptible to confounding factors including prompt design variations, reference construction methodology, and generation temperature settings, that remain outside the scope of our primary experimental controls.

\subsection{Key Findings and Insights}

Our comprehensive evaluation revealed several important insights:

\begin{itemize}
\item Naive chunking with prefix-fusion consistently delivered the strongest retrieval performance, achieving the highest MRR (0.750), NDCG (0.813), and precision (0.798), while matching the best Hit Rate@10 (0.900). Uniform chunk sizes ensure that prefixed metadata exerts consistent influence across all segments, whereas variable-length chunks from semantic or recursive strategies can unevenly distribute this influence—larger chunks may bury the prefix signal, while smaller ones overemphasize it. For formally structured documentation such as AWS S3~\cite{AWSS3Docs}, which features clear sectional boundaries and hierarchical organization, naive segmentation avoids the over-fragmentation that semantic methods introduce in dense technical text with varying information density.

\item Recursive chunking demonstrated robust performance across all embedding techniques, with content-only embeddings achieving competitive results (NDCG 0.782, Precision 0.783, MRR 0.713). Its hierarchical splitting preserves document structure at natural boundaries, providing reliable representations regardless of embedding methodology. This stability makes recursive chunking a dependable default when corpus structural properties are unknown or heterogeneous.

\item The weight sensitivity analysis reveals that metadata integration exhibits a seasoning effect: a 10\% metadata contribution consistently enhances retrieval across all chunking strategies, while higher ratios progressively degrade performance. This occurs because TF-IDF metadata vectors are inherently sparser and less semantically rich than content embeddings; excessive weighting amplifies shallow keyword matches over deep content relevance. Prefix-fusion sidesteps this limitation entirely by allowing the encoder to contextually modulate metadata influence during encoding rather than through post-hoc vector arithmetic.
\end{itemize}

These results empirically validate that metadata enrichment frameworks must be carefully tailored to specific retrieval objectives. The findings establish quantitative benchmarks for implementing high-performance RAG systems, highlighting the non-trivial interactions between chunking strategies and embedding techniques in determining retrieval effectiveness across standardized information retrieval metrics.

Table~\ref{tab:tradeoff} summarizes these findings as a configuration decision matrix for practitioners deploying metadata-enriched RAG systems in production environments.

\begin{table}[!t]
\centering
\caption{Configuration decision matrix for practitioners.}
\label{tab:tradeoff}
\setlength{\tabcolsep}{3pt}
\renewcommand{\arraystretch}{1.2}
\begin{tabular}{lll}
\hline
\textbf{Priority} & \textbf{Configuration} & \textbf{Key Metric} \\
\hline
Best Ranking   & Naive + Prefix-Fusion  & NDCG 0.813 \\
Max Hit Rate   & Naive + TF-IDF/PF      & HR@10 0.900 \\
Best Stability & Recursive + Content    & Prec. 0.783 \\
Lowest Latency & Content-Only           & P50 $\sim$12\,ms \\
Balanced       & TF-IDF ($w_c{=}0.9$)  & NDCG 0.687 \\
\hline
\end{tabular}
\end{table}

\section{Conclusion}
This study demonstrates that integrating LLM-based metadata enrichment into retrieval pipelines significantly enhances RAG system performance in technical documentation contexts. Our $3 \times 3$ factorial evaluation reveals that the method of metadata integration matters as much as the choice of chunking strategy. Prefix-fusion---which allows the encoder to jointly process metadata and content as a unified input---consistently outperforms TF-IDF weighting, where separately computed vectors are blended via linear combination under an assumption of vector space alignment that does not hold in practice. Naive chunking with prefix-fusion emerges as the strongest configuration on our corpus, as the uniform segment sizes ensure consistent metadata influence across chunks. This advantage is amplified by the inherently well-structured nature of AWS S3 documentation~\cite{AWSS3Docs}, where clear sectional boundaries reduce the value of semantically-aware splitting. Chunking strategy selection is thus fundamentally domain-dependent: formally structured corpora may favor simpler segmentation augmented with metadata, whereas less structured documents may benefit from adaptive semantic chunking. Recursive chunking offers the most stable cross-configuration performance, making it a robust default when corpus properties are unknown.

Metadata operates as a seasoning effect: a small proportion (10\%) consistently improves retrieval by injecting discriminative signals, while higher ratios degrade performance as sparse metadata features dilute the content representation. Prefix-fusion avoids this degradation entirely by enabling the encoder to contextually modulate metadata influence during encoding rather than through post-hoc vector arithmetic. These insights, combined with sub-30\,ms P95 latency and improved vector space organization (DBI: 3.12 vs.\ 3.77), confirm that metadata-enriched RAG systems are production-viable and provide actionable configuration guidance for practitioners.

Future work should explore scalability to larger and more diverse corpora, dynamic metadata generation for evolving content, and integration within real-time retrieval systems with adaptive chunking strategies.

\subsection{Limitations}
Our evaluation relies on silver-standard pseudo-ground truth derived from cross-encoder reranking rather than human annotations, and scores only chunks retrieved by at least one configuration, introducing retrieval pool bias. Consequently, absolute metric values should be interpreted with caution, while relative comparisons between configurations remain valid. Additionally, the evaluation is conducted on a single domain (AWS S3 documentation); replication on heterogeneous corpora is necessary to establish broader generalizability.

\section*{ACKNOWLEDGMENT}

This project has been completed under the supervision of Fatemeh Sarayloo. The first four authors were pursuing their master's programs while working on this project. We appreciate the support of the University of Illinois Business College.

\end{document}